\documentclass[twoside,reqno]{qts-proc}
\usepackage{epsfig,cite}
\usepackage{amssymb,amsmath}
\usepackage{times}
\usepackage{color}
\usepackage{graphics}
\usepackage{amsfonts}
\usepackage{shadow}
\newcommand{\beq}{\begin{equation}}
\newcommand{\eeq}{\end{equation}}
\newcommand{\beqa}{\begin{eqnarray}}
\newcommand{\eeqa}{\end{eqnarray}}
\newcommand{\nn}{\nonumber \\}

\newcommand {\np}[1]{ {\mbox{\textrm{:}}{#1}{\textrm{:}}} }

\def \mod { \, \mathrm{mod} \, }

\def \Im {\mathrm{Im}\ }
\def \df {\stackrel{\mathrm{def}}{=} \ }
\def \e {\mathrm{e}}
\def \el {\mathrm{el}}

\def \H {{\mathcal H}}
\def \uu {\widehat{u(1)}}

\def \D {\Delta}
\def \eps {\varepsilon}
\def \l {\lambda}
\def \Lu {\underline{\Lambda}}

\def \s {\sigma}

\def \t {\tau}
\def \z {\zeta}
\def \la {\langle}
\def \ra {\rangle}

\def \R {{\mathbb R}}

\def \Z {{\mathbb Z}}

\def \PF {\mathrm{PF}}
\def \ch {\mathrm{ch}}

\def \qh {\mathrm{qh}}

\begin{document}
\sloppy \raggedbottom
\setcounter{page}{1}

\newpage
\setcounter{figure}{0}
\setcounter{equation}{0}
\setcounter{footnote}{0}
\setcounter{table}{0}
\setcounter{section}{0}

\begin{start}

\title{Conformal field theory description of mesoscopic phenomena
in the fractional quantum Hall effect}

\runningheads{L. Georgiev}{CFT description of mesoscopic phenomena in the
FQH effect}

\author{Lachezar Georgiev}{1}

\address{Institute for Nuclear Research and Nuclear Energy \\
72 Tsarigradsko Chaussee, 1784-Sofia, Bulgaria}{1}

\begin{Abstract}
We give a universal description of the mesoscopic effects occurring in
fractional quantum Hall disks due to the Aharonov--Bohm flux threading
the system. The analysis is based on the exact treatment of the flux
within the conformal field theory framework and is relevant for all
fractional quantum Hall states whose edge states CFTs are known. As an
example we apply this scheme for the parafermion Hall states and
extract the main characteristics of the low- and high-
temperature asymptotic behavior of the persistent currents.
\end{Abstract}

\end{start}

\section{Introduction}
Mesoscopic systems are characterized by their intermediate size
in between the micro- and macro- systems---small enough so that
the electrons  still move in a  coherent way, yet big enough for
some measurable consequences such as the Aharonov--Bohm (AB) effect
to be  observable.
One important property of the mesoscopic rings threaded by AB flux is that
the free energy of the ring is a periodic
function of the flux with period one flux quantum $h/e$ (Bloch theorem).
In addition to its natural relevance for studying quantum effects
 mesoscopic physics has been at the core of the very promising
recent proposal \cite{sarma}
for topological quantum computers in terms of the braid matrices of
 non-abelian fractional quantum Hall (FQH) anyons.
The advantage of such topological gates, as compared to the other
 existing quantum computation schemes, is that
quantum information is encoded  in a non-local way
which makes it inaccessible to local interactions and decoherence.
In this talk we demonstrate that it is possible and very convenient
to compute  mesoscopic
quantities such as the persistent currents for FQH rings  purely
from their edge states conformal field theory (CFT) by using
the CFT partition function with flux.
The effect of adding AB flux is simply twisting the electron and
quasiparticle operators. The numerical results
\cite{NPB2001,PRB-PF_k} show that the period of the persistent currents
for the parafermion FQH states  \cite{rr,NPB2001,PRB-PF_k} is exactly
one flux quantum which  means that there could be
no spontaneous breaking of continuous symmetries.
We derive an analytic formula for the low-temperature
amplitudes of persistent currents which are shown to decay
logarithmically  due to thermal activation of q.h.--q.p. pairs.
Also we find an analytic formula for the  high-temperature
asymptotics of the persistent currents which decay exponentially
 due to thermal decoherence with universal  non-Fermi liquid exponent
derived from the CFT. These exponents can be used to characterize
unambiguously the FQH universality classes.
\section{Persistent currents: an intuitive picture}
\begin{center}
\begin{figure}[htb]
\caption{Corbino ring geometry for FQH system threaded by AB flux
\label{fig:corbino}}
 \includegraphics*[bb=150 320 470 600,clip,width=9cm]{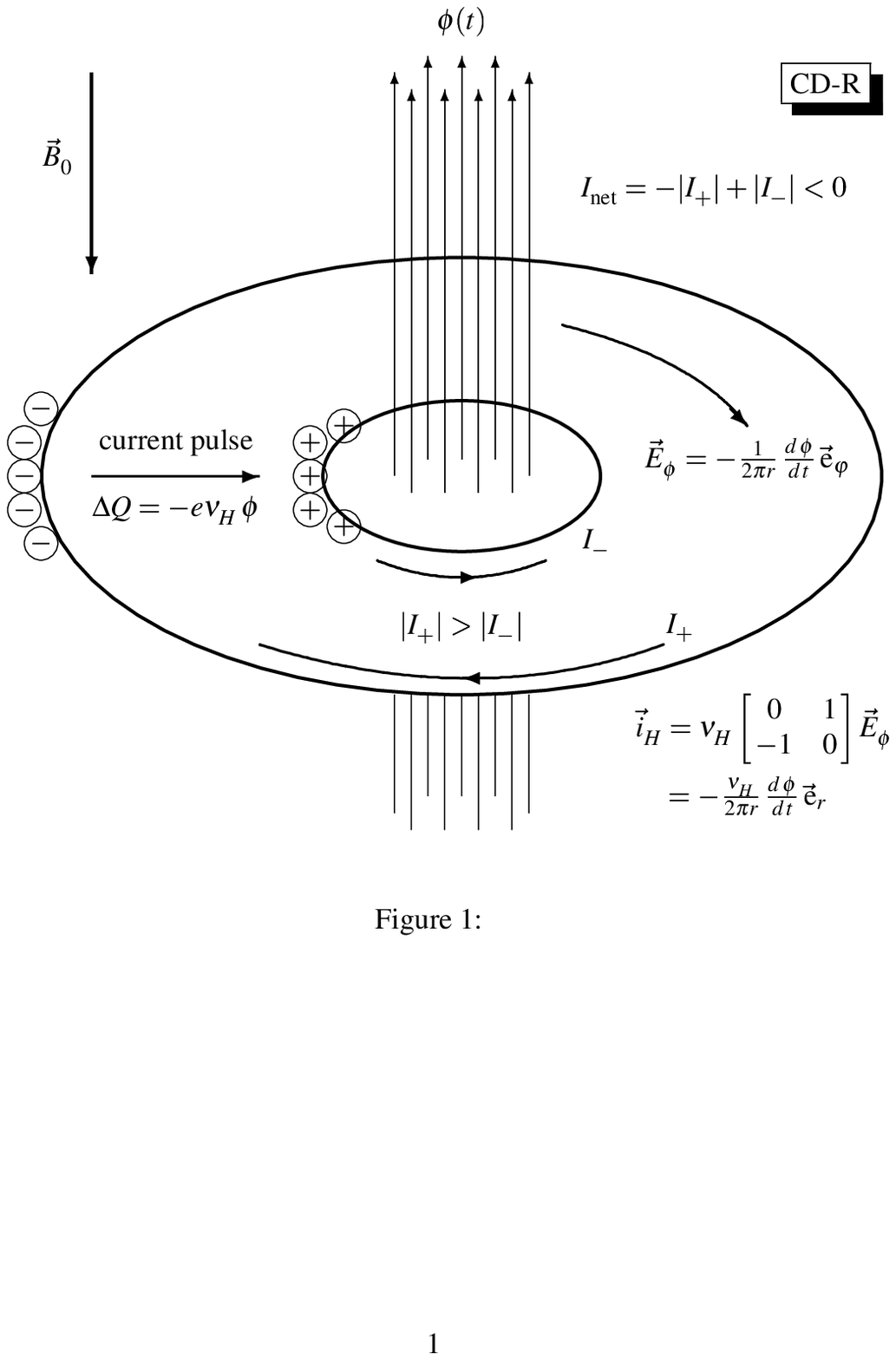}
\end{figure}
\end{center}

When at zero temperature the AB flux,  threading a FQH system in the
Corbino ring geometry shown on Fig.~\ref{fig:corbino},
 is increased adiabatically
this induces azimuthal electric field $\vec{E}_\phi$
however there is no azimuthal bulk current since $\s_{xx}=0$.
But because $\s_{H}\neq 0$ there is a radial current pulse which
transfers charges between the two edges.
This leads to edge currents imbalance and produces a non-zero net
current.
Despite the huge non-mesoscopic contribution \cite{oscillate}
these oscillating mesoscopic persistent currents could be measured
by two-point-contact SQUID detectors.
\section{The quantum Hall effect and conformal field theory}
In the classical Hall effect the conductance of
a two-dimensional conducting plate in
presence of normal magnetic field $B$ has a nonzero off-diagonal component
$\sigma_{H}$. The quantum Hall effect is characterized by the
quantized $\sigma_{H}$ and at the same time the vanishing diagonal conductance
\[
\widehat{\sigma}=
\left[ \begin{array}{cc} 0 & -\sigma_H \\ \sigma_H & 0  \end{array}\right],
\quad
{\sigma}_H=\left(\frac{e^2}{h}\right) \nu,\quad
\nu_H=\frac{n}{n_B}, \quad n_B=\frac{B}{h/e},
\]
where $n$ is the electron density and $\nu_H=n_H/d_H$ the filling factor.
In order to observe this effect one needs
high magnetic fields ($B\sim 1-30$ T), low temperature
($T < 1$ K), low electric fields, low electron density
($n\sim 10^{11} \ cm^{-2}$), high mobility
($\mu\sim 10^5 - 10^7 \ cm^2/V.s$) and extremely high-quality samples.
\subsection{Effective quantum field theory approach: the 2+1 dimensional
 Chern--Simons model}
The presence of the bulk energy gap in the low-energy spectrum of FQH
liquid implies that it is incompressible, i.e., when the system is on a
plateau of $\s_H$ changing $B$ slowly leads to low-energy (de)compression
which are suppressed by the gap and therefore $\s_H$ does not change.
The effective quantum field theory (QFT)  for the low-energy excitations
of the incompressible electron fluid
has been obtained in the thermodynamic scaling limit
\cite{fro-RMP,fro2000}---the non-relativistic interacting
Schr\"odinger electrons
plus incompressibility leads to  abelian $2+1$ dimensional
Chern--Simons QFT
\[
S_{\mathrm{eff}}=\frac{\sigma_H}{2}\int  a \wedge d a, \quad
(a_\mu)=\frac{e}{\hbar} (A_0,-A_1, -A_2).
\]
The current in response to the external electromagnetic field
for a finite sample with boundaries
\[
j^\mu=\frac{\delta S_{\mathrm{eff}}(A)}{\delta A_\mu}=
\sigma_H \varepsilon^{\mu\nu\rho} \partial_{\nu} A_\rho
\quad\Longrightarrow \quad
\left\{ \begin{array}{l} j_0=\sigma_H \, B \\
j_k=\sigma_H \, \varepsilon_{kl}\, E_l \end{array}\right.
\]
turns out to be anomalous. This induces a current $j_\alpha$
along the boundaries
\[
\partial_\alpha j^\alpha=\frac{1}{2}\sigma_H \epsilon_{\alpha\beta}
F_{\alpha\beta},
\quad \sigma_H=\frac{e^2}{h} \nu_H, \quad
F_{\alpha\beta}=\partial_\alpha A_\beta - \partial_\beta A_\alpha,
\]
which is anomalous too, however, the bulk anomaly is exactly
compensated by that of the edge currents so that the total current
is conserved. This bulk--edge anomaly cancellation
is the origin of the so called holographic principle
implying the  precise correspondence between bulk and edge spectra
\cite{fro2000}.
\subsection{The correspondence between Chern--Simons and RCFT}
According to  Witten's correspondence
the $(2+1)$ dimensional  Chern--Simons QFT on the dense cylinder
$(t,x)\in \R\times\mathrm{Disk}$ is equivalent to
the  rational $(1+1)$ dimensional  CFT on the spacetime  border
$(t,\tilde{x})\in \R\times\mathrm{Circle}$.
This correspondence is not exactly one-to-one:
one  bulk  QFT corresponds to many boundary CFTs, however
one  boundary  CFT corresponds to only one bulk QFT.
The correlation functions of the bulk observables
can be obtained from the  CFT correlation functions
by  analytic continuation.
Because the edge states dynamics determines all universal properties
it is believed that the  FQH universality classes can be labeled by
the rational CFT  (RCFT) for the edge states.
\subsection{CFT description of quasiparticles}
\label{sect:RCFT}
The CFT for the FQH edges always contains
$\uu\times Vir$ symmetry. The first factor describes
the electric charge while the second one---the angular momentum.
The FQH universality class can be described in terms of a single-edge
chiral CFT, i.e., corresponding to disk FQH states.
The quasiparticle excitations are labeled by the weights of the
irreducible representations of the CFT chiral algebra \cite{fro2000,NPB-PF_k}.
One of the most important characteristics of
this CFT is its {\it topological order}---the number of
topologically inequivalent quasiparticles whose irreducible
representation spaces are closed under fusion---which
in the CFT language is equal to the dimension of the modular $S$ matrix.
In addition the RCFT for a FQH ring with two edges must satisfy 4
invariance conditions ($T^2$, $S$, $U$, $V$) \cite{cz}.

The electron and the basic quasihole in the CFT are represented by
the primary fields \cite{fro2000,NPB-PF_k}
\[
\psi_\el(z)=\np{\e^{ -i\frac{1}{\sqrt{\nu_H}}\phi^{(c)}(z)}}
\otimes \psi^{(0)}(z),\quad
\psi_\mathrm{qh}(\eta)=
\np{\e^{ i\frac{1}{\sqrt{n_Hd_H}}\phi^{(c)}(\eta)}}\otimes
\psi^{(0)}_{\mathrm{qh}}(\eta).
\]
where $\psi^{(0)}(z) \in Vir$ and $\psi^{(0)}_{\mathrm{qh}}(\eta)\in Vir$
are the neutral component of the electron and the quasihole, respectively.
The wave functions of  $N$ electrons and $k$ quasiholes
(ground state corresponds to $k=0$) are computed as chiral
CFT correlation functions
\[
\Psi(\eta_1,\ldots, \eta_k;z_1,\ldots,z_N)=\la 0|\psi_\qh(\eta_1)
\cdots\psi_\qh(\eta_k)\psi_\el(z_1)\cdots \psi_\el(z_N) |N,k\ra .
\]
Finally, we would like to introduce the notion of
\textit{mesoscopic FQH rings}. We consider a single edge FQH system
on a Corbino disk. The bulk energy gap implies that
all low-energy excitations are living on the edge
which, under certain experimental conditions
($R\sim 5 \ \mu$m, $T\sim 1$ mK)  can be considered  a
mesoscopic ring displaying Aharonov--Bohm oscillations of the
magnetization \cite{sivan-imry}.
While the disk geometry is a little inconvenient for analyzing
flux-threading phenomena, the chiral CFT description presented here
appears very useful in antidot FQH states threaded by
Aharonov--Bohm flux \cite{geller-loss,5-2AB}.
\section{Aharonov--Bohm flux: twisting the electrons}
In this section we shall consider the effect of adding Aharonov--Bohm
magnetic field described by the vector potential
\[
\mathbf{A}=\left(\frac{h}{e}\right)\frac{\phi}{2\pi r^2}\, (-y,x) =
\left(\frac{h}{e}\right)\frac{\phi}{2\pi r} \ \mathbf{e_{\varphi}}, \quad
\phi \in \R \quad
\left(\mathrm{rot}\,\mathbf{A}=\phi \ \delta^{(2)}(\mathbf{r}) \right)
\]
where $\phi$ is the  AB flux in units $\phi_0=h/e$,  $r=\sqrt{x^2+y^2}$,
to quantum Hall systems \cite{NPB-PF_k}.
Because the AB magnetic field is zero along the edge, the electron operator
in AB field could be obtained by explicit computation \cite{NPB-PF_k} of the
line integral below
\beq\label{el_AB}
\psi^A_\el(z)=\exp\left(-i\frac{e}{\hbar} \int\limits_{*}^{z}
{\bf A} . d {\bf  r}  \right)
\ \psi_\el(z)=
z^{-\phi}\ \psi_\el(z),  \quad z=x+iy,
\eeq
where $\psi_\el(z)$ is the electron in the absence of flux.
This procedure is known in CFT as  orbifold twisting.
Note that the AB flux twists only the $\uu$ part of the electron
\cite{NPB-PF_k}
which is the usual vertex exponent of a normalized $\uu$ boson
\[
\np{\e^{ i\alpha \phi^{(c)}(z)}} \ \df
U_\alpha\, \e^{i\alpha \phi^{(c)}_+(z)} \, z^{\alpha J_0} \,
\e^{i\alpha \phi^{(c)}_-(z)},
\quad
i\phi^{(c)}_\pm(z)=\pm \sum_{n=1}^\infty J_{\mp n}\frac{z^{\pm n}}{n},
\]
where $U_\alpha$ are the (outer) charge-shift  automorphisms of the $\uu$
current algebra
generated by the Laurent modes of the  current
$J(z)=i\,  \partial \phi^{(c)}(z)=\sum_{n\in\Z} J_n z^{-n-1}$
with commutation relations
 $ \left[J_n,J_m\right]=n \, \delta_{n+m,0}$, i.e.,
\[
\left[J_n,U_\alpha\right]=\alpha \, U_\alpha \, \delta_{n,0}, \quad
U_\alpha\, U_\beta=U_{\alpha+\beta}, \quad \alpha,\beta\in\R,
\quad
U_0=1, \quad \left(U_\alpha\right)^\dagger= U_{-\alpha} .
\]
Comparing the twisting results \cite{kt}
for a  general twist  $\pi_\beta\ : \ A \to  U_{\beta}\, A \, U_{-\beta}$
\[
\np{\e^{ i\alpha \phi^{(c)}(z)}} \quad
\stackrel{\pi_\beta}{\longrightarrow} \quad
U_{\beta} \, \np{\e^{ i\alpha \phi^{(c)}(z)}} \, U_{-\beta} =
z^{-\alpha\beta}\,
\np{\e^{ i\alpha \phi^{(c)}(z)}}     ,
\]
where $\alpha= -1/\sqrt{\nu_H}$ for  the electron,
with $\nu_H$ the filling factor,
to Eq.~(\ref{el_AB}) we find that the twist $\beta$  is proportional
to the AB flux  $\phi$
\[
- \left(\frac{-1}{\sqrt{\nu_H}}\right)\, \beta =
-\phi\quad \Longrightarrow \quad
\beta \equiv -\sqrt{\nu_H}\, \phi .
\]
The electric $\uu$ current $J^{\mathrm{el}}(z)=i\sqrt{\nu_H} \,
\partial \phi^{(c)}(z)=\sum_{n\in \Z} J^{\mathrm{el}}_n \ z^{-n-1}$
becomes twisted as well \cite{kt}
\beq\label{J-AB}
J^{\mathrm{el}}(z) \quad \stackrel{\pi_\beta}{\longrightarrow} \quad
J^{\mathrm{el}}(z)+\frac{\nu_H\phi}{z} \qquad
\Longleftrightarrow \qquad
J^{\mathrm{el}}_n \quad  \stackrel{\pi_\beta}{\longrightarrow} \quad
J^{\mathrm{el}}_n+\nu_H\phi \ \delta_{n,0}.
\eeq
so that the  $\uu$  part of the (Sugawara) stress tensor
$T^{(c)}(z)=(1/2)\np{J(z)^2}$ is twisted too
\[
L_n^{(c)} \  \stackrel{\pi_\beta}{\longrightarrow} \
L_n^{(c)}+\phi J^{\mathrm{el}}_n +  \nu_H{\phi^2 \over 2} \delta_{n,0}.
\]
Thus the total stress tensor
$T(z)= T^{(c)}(z) + T^{(0)}(z)=\sum_{n} \ L_n \ z^{-n-2} $
is modified by the AB flux \cite{NPB-PF_k} as follows
\beq\label{T-AB}
 L_n = L_n^{(c)}+ L_n^{(0)} , \qquad   \qquad
L_n \  \stackrel{\pi_\beta}{\longrightarrow} \
L_n+\phi J^{\mathrm{el}}_n +  \nu_H\frac{\phi^2}{2} \delta_{n,0}.
\eeq
The  chiral CFT partition function which we shall use as a
thermodynamic potential for $T\geq 0$ is computed as the trace of
the Boltzmann factor over the Hilbert space,
$\H=\mathop{\oplus}_{\lambda=1}^N \H_\lambda$, containing all
superselection sectors  corresponding to all kind of static
topologically inequivalent point-like sources  within the disk,
\[
Z(\t,\z)=
\mathop{\mathrm{tr}}_{\quad  \H \ }
\e^{2\pi i\left[\t\left(L_0 -\frac{c}{24}\right) +
\z J_0^\mathrm{el}\right]} = \sum_{\l=1}^N \chi_{\l}(\t,\z),
\]
Now that we know  the expressions for the stress
tensor (\ref{T-AB}) and electric current  (\ref{J-AB})
in presence of AB flux we can compute
the total chiral CFT partition function in  presence of AB flux $\phi$
\beqa\label{Z_phi}
Z_{\phi}(\t,\z) &=& \mathop{\mathrm{tr}}_{\quad  \H \ }
\e^{2\pi i\left[\t\left(L_0 +\phi J_0^\mathrm{el} +
\nu_H \frac{\phi^2}{2} -\frac{c}{24}\right) +\z \left(J_0^\mathrm{el} +
\nu_H\phi\right)\right] } \nn
&=&  \e^{2\pi i \nu_H\left[ \frac{\phi^2 }{2}\t +\phi\z\right] }\
Z(\t,\z+\phi\t).
\eeqa
Equation (\ref{Z_phi}) is our main result in this section:
adding AB flux $\phi$ essentially amounts to shifting  the modular
parameter
$\z\to\z+\phi\t$. The precise thermodynamical identification of the
modular parameters, which for a chiral sample (single circular edge of
circumference $L$ and edge velocity $v_F$) have to be pure imaginary,
with the absolute temperature $T$ and AB flux $\phi$ is
\cite{NPB-PF_k}
\[
q=\e^{2\pi i \t}, \quad \t=i\pi \frac{T_0}{T},\quad \z=\phi\t, \quad
T_0=\frac{\hbar v_F}{\pi k_B L}, \quad \phi\in \R.
\]
\section{Persistent currents for the  $\Z_k$ parafermion FQH states}
\subsection{The RCFT for the $\Z_k$ parafermion FQH states }
The RCFT for the $\Z_k$ parafermion FQH states
has  the general $\uu\times Vir$ structure emphasized in
Sect.~\ref{sect:RCFT}
\beq\label{RCFT}
\left(\widehat{u(1)} \oplus \PF_k \right)^{\Z_k},
\quad  \PF_k=\frac{\widehat{su(k)_1}\oplus\widehat{su(k)_1}}{\widehat{su(k)_2}}
\eeq
where the neutral $Vir$ component of the chiral algebra is realized
as a diagonal affine coset \cite{NPB2001,NPB-PF_k,diag_coset}.
The FQH states corresponding to the RCFT in (\ref{RCFT}), which
correspond to
central charge $c= 1 + 2(k-1)/(k+1)$  and
 filling factor $\nu_k= 2 + k/(k+2)$
have been introduced in \cite{rr}. The edge excitations are represented
 by the primary fields of (\ref{RCFT}) labeled by
$(\lambda,\Phi) \Rightarrow  \
:\e^{i \frac{\lambda}{\sqrt{k(k+2)}}\phi^{(c)}(z) }:  \otimes \
\Phi(z)$, where the neutral excitation is a $\Z_k$ parafermion primary field
$\Phi(z) \in \PF_k^*$.
The superscript $\Z_k$ denotes a selection rule called
the \textit{$\Z_k$ Pairing Rule} (PR) \cite{NPB2001,NPB-PF_k}
which says that an excitation with label $(\lambda,\Phi)$
is allowed only if
\[
P\left[ \Phi\right] = \lambda \ \mod \ k ,
\]
with $P\left[ \Phi\right]$ the parafermion $\Z_k$-charge of $\Phi$.
For more details about the parafermion FQH states and the diagonal
 coset construction see Refs.~\cite{rr,NPB2001,NPB-PF_k,diag_coset}.
\subsection{Partition function and persistent currents}
According to the standard thermodynamic interpretation  the partition
function computed in the CFT is related to the
 free energy by $F(T,\phi)=-k_B T \ln Z(T,\phi)$ and its dependence on
the AB flux gives rise to an equilibrium persistent current
\[
I(T,\phi)=\left(\frac{e}{h}\right)\, k_B T \frac{\partial}{\partial \phi}
\ln Z(\t,\phi\t).
\]
The total chiral partition function for the  $\Z_k$ parafermion states
is
\[
Z_{k}(\t,\z) =
\e^{-{\pi}\nu_H\frac{\left(\Im\z\right)^2}{\Im\t}}
\sum_{l  \mod  k+2} \quad
\sum_{\rho \geq l-\rho \mod k } \chi_{l,\rho}(\t,\z),
\]
where the \textit{characters} of the RCFT (\ref{RCFT})
are expressed in terms of the $\uu$ characters
$K_{l}(\t,k\z;k(k+2)) $ and  those,
$\ch\left(\Lu_{l-\rho+s} +\Lu_{\rho+s}\right)(\t)$,  for the parafermions
\[
\chi_{l,\rho}(\t,\z) =
\sum_{s=0}^{k-1} K_{l+s(k+2)}(\t,k\z;k(k+2))
\ch\left(\Lu_{l-\rho+s} +\Lu_{\rho+s}\right)(\t),
\]
whose explicit forms can be found in
Refs.~\cite{NPB2001,NPB-PF_k,diag_coset}.
\subsection{Low temperature regime:  $T\ll T_0$}
Because in the low-temperature limit $T\to 0$ the modular parameter
$q=\exp\left(-2\pi^2 \frac{T_0}{T}\right) \to 0$
we can keep only the leading terms in $Z(\t,\z)$
which come from the RCFT sectors with lowest CFT-dimensions only, i.e.,
the vacuum and the one-quasiparticle and one-quasihole sectors
\[
Z(T,\phi)  \mathop{\simeq}\limits_{\ T\ll T_0 \ }
q^{\frac{\nu_H}{2}\phi^2} \left\{ 1+ q^{\D_\qh}
\left( q^{Q_\qh\phi}+q^{-Q_\qh\phi}\right)\right\},
  \quad Q_\qh =\frac{1}{k+2},
\]
from which we obtain the low-T asymptotics of persistent current amplitude
\beq\label{I_max}
I_{\max}(T)\  \mathop{\simeq}\limits_{T\ll T_0} \ \nu_H   \frac{e v_F}{L}
\left\{ \frac{1}{2} -\frac{k_B T}{2 \widetilde{\eps}_\qh}
  \left[1+\ln\left( \frac{1}{n_H}\frac{2 \widetilde{\eps}_\qh}{k_B T}\right)
\right]
	\right\},
\eeq
where $\widetilde{\eps}_{\qh}$ is the proper quasihole energy
\cite{PRB-PF_k,NPB-PF_k}
\[
\widetilde{\eps}_{\qh}= 2 \pi^2 k_B  T_0 \ \D_\qh, \qquad
 \D_\qh=\frac{1}{2n_H d_H} +\D^{(0)}_\qh ,\qquad
\D^{(0)}_\qh= \frac{k-1}{2k(k+2)}.
 \]
The low-temperature logarithmic decay of the persistent currents amplitudes
(\ref{I_max})  can be interpreted as  due to thermal activation of
quasiparticle--quasihole pairs.
Note that the fundamental quasiholes in the parafermion FQH states
 have the  lowest electric charge $1/(k+2)$  and CFT dimension $1/(2k+4)$.
\subsection{High temperature regime: $T\gg T_0$}
At very high temperatures the modular parameter
$q=\exp\left(-2\pi^2 \frac{T_0}{T}\right) \to 1$
in which limit  the partition function is divergent.
Therefore, it is convenient to first perform a modular $S$ transformation
$(\t,\z)\to (-1/\t,-\z/\t)$
after which the new modular parameter
$q'=\exp(2\pi i \t')\to 0$ for $T\to\infty$.
Now the high-temperature expansion of the partition function
is determined by the leading terms, however for $q'\to 0$ in the characters
$\chi_{\l}(\t',\z')$,
\[
Z(\t,\z)= \sum_{\l=1}^N \chi_{\l}(\t',\z')
\left(\sum_{\l'=1}^N S_{\l\l'}\right)
\]
for which the $S$ matrix is  explicitly needed.
Using the $S$ matrix computed in \cite{NPB-PF_k}
we obtain the following exponential decay of the persistent
currents amplitudes
\[
\overline{I}_{k}(T) \ \mathop{\simeq}_{T\gg T_0} \ I^0_k \left(\frac{T}{T_0}\right)
\exp\left(-\alpha_k \frac{T}{T_0}\right),
\quad I^0_k=\mathrm{const.}
\]
\begin{figure}[htb]
\centering
\epsfig{file=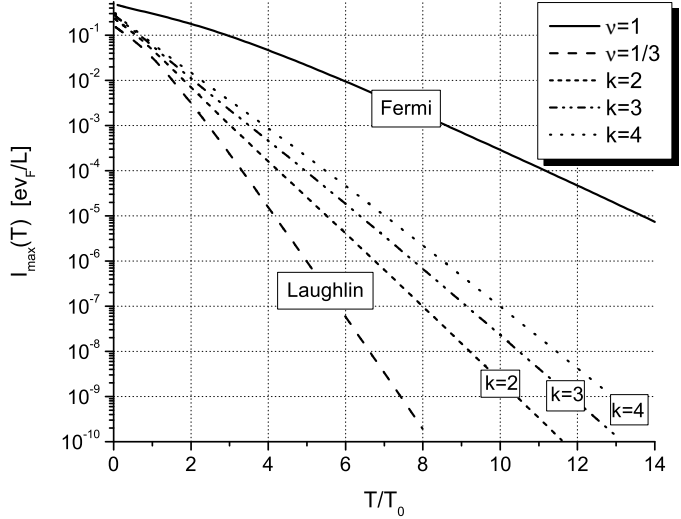,width=10cm}
\vspace{-0.5cm}
\caption{Logarithmic plot of the temperature decay of the persistent
current's amplitudes for the Fermi liquid ($\nu_H=1$), the
Laughlin state ($\nu_H=1/3$) and the $k=2,3$ and $4$ parafermion FQH
states \label{fig:log-decay}}
\end{figure}
This universal thermal decoherence is described by the
universal exponents $\alpha_k$, which
determine the slopes of the logarithmic plots in
Fig.~\ref{fig:log-decay}. Notice that,
unlike the Fermi/Luttinger liquids  where $\alpha=\nu^{-1}$,
$\alpha_k$ have a crucial neutral contribution
\beq\label{alpha}
 \alpha_k = \frac{1}{\nu_H}+
2\D^{\PF_k}\left(\Lu_0+\Lu_2\right)=
\frac{k+6}{k+2},
\eeq
where $\D^{\PF_k}\left(\Lu_0+\Lu_2\right)=(k-2)/(k(k+2))$ is the neutral
CFT dimension of the parafermion primary field with label
$\Lu_0+\Lu_2$.
Thus the universal decay exponents (\ref{alpha})
 can be considered  the fingerprint of the FQH states.
\section*{Acknowledgments}
This work has been partially supported by the Bulgarian
National Council for Scientific Research under Contract F-1406
and by the FP5-EUCLID Network Program  of the European
Commission under Contract HPRN-CT-2002-00325.
\bibliography{Z_k,my,qc}

\end{document}